\documentclass[preprint,showpacs,preprintnumbers,amsmath,amssymb]{revtex4}

\usepackage{graphicx}
\usepackage{dcolumn}
\usepackage{bm}
\usepackage{amsmath}
\usepackage{amssymb}
\usepackage{subfigure}
\usepackage{psfig,float}
\def\OMIT#1{#1}
\begin{document}

\title{Clustering under the line graph transformation: Application to reaction network}

\author{J. C. Nacher$^*$, N. Ueda, T. Yamada, M. Kanehisa, and T. Akutsu.}
\affiliation{Bioinformatics Center, Institute for Chemical Research, Kyoto University, 
Uji, 611-0011, Japan}

\date{\today}





\begin{abstract}

Many real networks can be understood as two complementary networks with two kind of nodes. This is the case of metabolic networks 
where the first network has chemical compounds as nodes and the second one has nodes as reactions. The second 
network can be related to the first one by a technique called line graph transformation (i.e., edges in an initial network are 
transformed into nodes). Recently, the main topological properties of the metabolic networks have been properly described
by means of a hierarchical model. In our work, we apply the line graph transformation 
to a hierarchical network and the clustering coefficient $C(k)$ is calculated 
for the transformed network, where $k$ is the node degree. While 
$C(k)$ follows the scaling law $C(k)\sim k^{-1.1}$ for the initial 
hierarchical network, $C(k)$ scales weakly as $k^{0.08}$ for the transformed network. These results indicate that 
the reaction network can be identified as a degree-independent clustering network.

\end{abstract}

\maketitle


\section{ Introduction}

Recent studies on network science demonstrate that cellular networks 
are described by universal features, which are also present in non-biological complex systems, as for example social networks or 
{\it WWW} \cite{review}. Most networks encountered in real world have scale-free topology, in particular networks of fundamental 
elements of cells 
as proteins and chemical substrates \cite{wuchty, wagner, jeong}. In 
these networks, the distribution 
of node degree follows a power-law as $P(k)\sim k^{-\gamma}$ (i.e., frequency of the nodes 
that are connected to $k$ other nodes). The degree of a node is the number of other nodes 
to which it is connected.

One of the most successful models for explaining that scale-free topology was proposed 
by {\it Barab\'{a}si-Albert}
\cite{bar1,bar2}, which introduced a mean-field method to simulate the growth
dynamics of individual nodes in a {\it continuum theory} framework. However, although that
 model was a milestone 
to understand the behavior
of real complex networks, it could not reproduce all the observed features in real networks such
as clustering dependence.
In order to bring under a single framework all the observed properties 
of biological networks  {\it
Rasvasz et al.} suggested successfully a hierarchical and modular
topology \cite{jerar1}. These observed properties of networks with $N$ nodes are: scale-free of degree 
distribution $P(k)\sim k^{-\gamma}$ , power-law scaling of clustering 
coefficient $C(k)$ and a high value for the average of the 
clustering coefficient $C(N)$ and its independence with network size. A network with these properties is called
hierarchical network.

In the hierarchical model \cite{jerar1} (the RSMOB model in what follows), the network is simultaneously 
scale-free and has a high clustering coefficient 
that is independent of the network size. The key signature of hierarchical modularity is the dependence of the clustering 
coefficient 
as a function of the node degree $k$, which follows
$C(k)\sim k^{-1}$. The meaning of this result is that nodes with a few links have a high clustering degree, being the 
centers of numerous interlinked
modules. On the other hand, highly connected nodes (hubs) have smaller clustering coefficient, being their tasks to 
connect different
modules. In \cite{jerar1, jerar2}, it is shown that many real networks (biological and non-biological)
have a hierarchical organization. One of them, which is the subject of our study, is the metabolic network.

It is also interesting to note 
that the metabolic network is an example of bipartite networks \cite{review,bi1,bi2}. In 
a bipartite network there are two kinds of nodes and edges only connect nodes of different kinds. In the metabolic network 
these nodes are chemical compounds and reactions. The network generated by the chemical compounds (reactions) is 
called compound (reaction) projection. A line graph transformation (i.e., each edge between two compounds 
becomes a node (reaction) of the transformed network) relates both projections. A detailed analysis of 
the line graph transformation focused on the degree distribution $P(k)$ and 
applied to the metabolic network can be found in \cite{nacher}. In that work, similarities 
and differences between the line graph transformation and the metabolic network are discussed. There it was found that
if the initial network follows a power-law $P(k)\sim k^{-\gamma}$, the transformed network preserves the scale-free topology 
and in most cases the exponent is increased by one unit as $P(k)\sim k^{-\gamma + 1}$.  

The observed topological properties related to the clustering degree of the metabolic 
network (in particular, the chemical compound network) have been properly described
by means of the RSMOB model. In the present work, our aim is to study the
clustering coefficients $C(k)$ and $C(N)$ of the reaction network by using two approaches: Firstly, 
we derive mathematical equations of those coefficients 
in the transformed network. Secondly, we apply 
the line graph transformation to a hierarchical network. The results from both methods are compared 
with experimental data of reactions from KEGG database \cite{KEGG} showing a good agreement. Though we started 
this work motivated by theoretical interest in the line graph transformation, the results provide explanation for the difference of $C(k)$
between the compound network and the reaction network.
 
In our work, the hierarchical network is generated by the RSMOB model, where 
the nodes correspond to chemical compounds and the edges
correspond to reactions. While the RSMOB model reproduces successfully the hierarchical properties
of the compound network, here we show that this hierarchical model also stores  adequate information 
to reproduce the experimental data of the
reaction network. Our study indicates that it is enough to apply the line graph transformation to the hierarchical network 
to extract that information. While $C(k)$ follows the power-law $k^{-1.1}$ for the initial hierarchical network (compound network), $C(k)$ scales weakly as $k^{0.08}$ 
for the transformed network (reaction network).
Consequently, we conclude that the
reaction network can not be defined as a hierarchical network.


It is also worth noting that the line graph transformation has recently been applied with success 
on the protein interaction network \cite{protein} with the aim to detect functional modules. In that work, the 
edges (interactions) between two proteins 
become the nodes of the transformed network (interaction network). By means of the line graph transformation, the interaction
network has its structure level more increased than that from the protein network (i.e., higher clustering coefficient). By 
using the TribeMCL algorithm \cite{protein2} they are able to detect clusters in the more highly 
clustered interaction network. These clusters are transformed back to the initial protein-protein network to identify 
which proteins conform functional clusters. At this point, we note that the aim of our study is not to detect functional modules from the
metabolic network. In our work the line graph transformation is used successfully to evoke topological properties 
related to the clustering degree of the reaction network. 

The paper is organised as follows. In Sec. II, we describe the theoretical concepts in our approach, and explain the mathematical methods
used in our analysis. In Sec. III we present the 
experimental data of metabolic pathways of the KEGG database for $C(k)$ and $C(N)$, and we compare with our
theoretical predictions before and after the line graph transformation is done. The final section summarizes our work.


 
\section{ THEORETICAL APPROACH}
\subsection{Clustering coefficients $C(k)$ and $C(N)$}

Recent analyses have demonstrated that the metabolic network has 
a hierarchical organization, with properties as: scale-free degree
distribution $P(k)\sim k^{-\gamma}$, power-law dependence of
clustering coefficient $C(k)\sim k^{-1}$ and independence with network size of the average 
clustering coefficient  $C(N)$, where $N$ is the total number of
nodes in a network \cite{jerar1}. The clustering coefficient 
can be defined for each node $i$ as: 

\begin{equation}
C_i(k_i) = \frac{2n_i}{k_i(k_i-1)}, 
\end{equation}
where $n_i$
denotes the number of edges connecting the $k_i$ nearest neighbors of node $i$ to each other. $C_i$ is equal to 1 for a 
node at the center of a fully
interlinked cluster, and it is 0 for a node that is a part of a loosely connected cluster \cite{jerar1}. An example 
can be seen in Fig. 1(a).
\OMIT{
\begin{figure}[h]
\centerline{\protect
\hbox{
\psfig{file=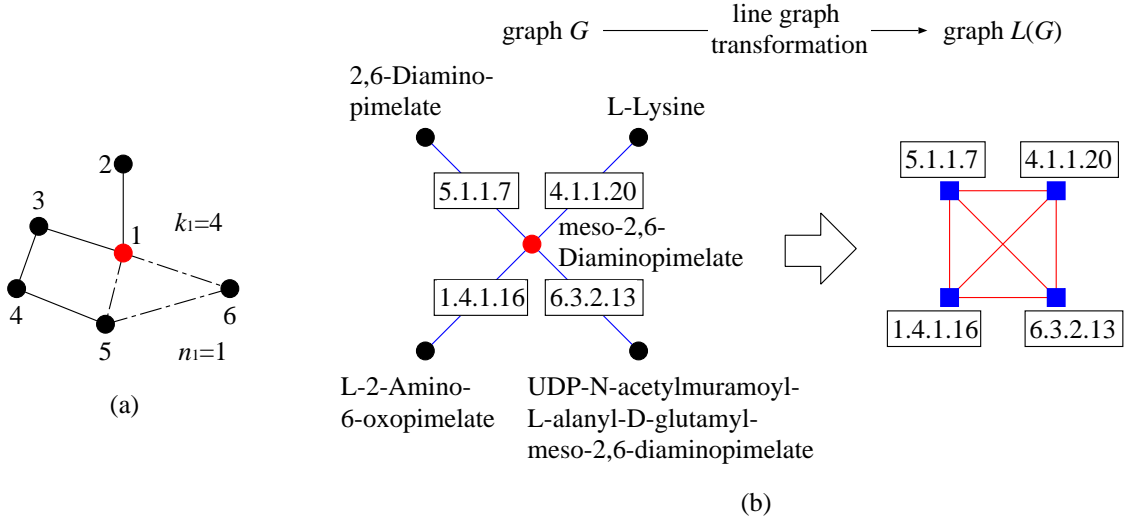,height=7.0cm,angle=0}}}
\caption{\small{a) Example of clustering in an undirected network. Continuous and dash-dotted lines mean interaction between nodes. In
addition, the dash-dotted line defines the only triangle where the node 1 (red) is one of the vertices. The node 1 has 4 neighbors
($k_i$ = 4), and among these neighbors only one pair is connected ($n_1$=1). The total number of possible triangles that could go 
through node $i$ is 6. Thus, the clustering coefficient has the value $C_1$ = 1/6. High density of triangles means high
clustering coefficient. b) We show an example of the line graph transformation. The initial graph $G$ corresponds 
to one subgraph
which belongs to the Lysine Biosynthesis metabolic pathway. This graph is constructed by taking nodes as chemical compounds and edges
as reactions. By applying the line graph transformation we find graph $L(G)$, which is the reaction graph embedded in the graph $G$. The
nodes of the graph L(G) are the reactions of the graph $G$ \cite{nacher}.}}
\end{figure}
}
Geometrically, $n_i$ gives the number of triangles
that go through node $i$. The factor $k_i(k_i-1)/2$ gives the total number
of triangles that could go through node $i$ (i.e., total number of triangles obtained when all the neighbors of node $i$ are 
connected to each other). In the case 
of Fig. 1(a), there is one triangle that 
contains node 1
(dash-dotted lines), and a total of 6 triangles 
could be generated as the maximum. Hence, the clustering 
coefficient of node 1 is $C_1$ = 1/6.

On the other hand, the average clustering 
coefficient $C(N)$ characterizes the overall tendency of nodes to form clusters as a function
of the total size of the network $N$. The mathematical expression is: 

\begin{center}
\begin{equation}
C(N)=\frac{1}{N} \sum_i C_i(k_i).
\end{equation}
\end{center}
 
The structure of the network is given by the function $C(k)$, which is defined as the average clustering coefficient over nodes with
the same node degree $k$. This function is written as:

\begin{equation}
C(k) = \frac{1}{N_k}\sum_{i:k_i=k}C_i(k_i) 
\end{equation}
where $N_k$ is the number of nodes with degree $k$, and the sum runs over the $N_k$ nodes with degree $k$. A scaling 
law $k^{-1}$ for this magnitude is an indication of the hierarchical topology of a network.

Once the theoretical definitions have been introduced, our aim is to analyse how the coefficients $C(N)$ and
$C(k)$ are modified under the line graph transformation.

\subsection{Line graph transformation on metabolic networks: spurious nodes}

Given an undirected graph $G$, defined by a set of nodes $V(G)$ and a set of edges $E(G)$, we associate another graph $L(G)$, called
the line graph of $G$, in which $V(L(G))=E(G)$, and two vertices are adjacent if and only if they have a common endpoint in $G$ (i.e.,
$E(L(G))=\{\{(u,v),(v,w)\}| (u,v) \in E(G), (v,w) \in E(G)\}$). This construction of
graph $L(G)$ from the initial graph $G$ is called line graph transformation \cite{15}.

It is worth noting that in a previous work \cite{nacher} the degree distribution $P(k)$ was studied 
by applying line graph transformation to synthetic and real
networks. There it is assumed an initial 
graph $G$ with scale-free topology as $P(k)\approx k^{-\gamma}$. 
As the degree of each transformed node (i.e., an edge in $G$) will be roughly around $k$, the distribution of the line graph $L(G)$
should be $k\cdot k^{-\gamma}$ = $k^{-\gamma + 1}$ with degree around $k$. Therefore, it is concluded that if we have a 
graph $G$ with a probability distribution following a power-law as $k^{-\gamma}$, then
$L(G)$ will follow a power-law as $k^{-\gamma + 1}$. The real networks under study were protein-protein
interaction, {\it WWW}, and metabolic networks. In Fig. 1(b), we can see an example of the line graph transformation applied 
to a subgraph of the
metabolic network.

However, it is important to point out one issue. In metabolic networks, there are cases where spurious nodes appear. For example, we consider two reactions sharing the same substrate (or product) and at least one of the chemical reaction
has more than one product (or substrate). If we apply a line graph transformation to this network, we would obtain more than two nodes in
the transformed network, where only two nodes (reflecting two reactions present in the real process) should appear. These spurious
nodes appear only when one (or some) reaction(s) in the network has 
more than one product (or substrate). Therefore, these cases should be computed and transformed by generating only as many nodes 
in the transformed network as reactions
in the real metabolic process. This procedure is called ${physical}$ line graph transformation. In the present work, we have 
applied this procedure 
to generate the reaction network by using experimental data from the KEGG database. Experimental data \cite{note1} 
are shown later in Figs. 6 and 7 
(blue diamonds). More detailed information about this issue can be found in \cite{nacher}. 



\subsection{Equations of $C(k)$ and $C(N)$ under the line graph transformation}

We assume a graph $G$ as it is depicted in Fig. 2(a). In this graph, edge $a$ connects two nodes with 
degree $k^\prime$ and $k^{\prime\prime}$. We apply the
line graph transformation to this graph $G$ and the result of this transformation is the line graph of $G$, $L(G)$ shown in Fig. 2(b).
We see that, under the line graph transformation, the nodes of $L(G)$ are the edges of $G$, with two nodes of $L(G)$ adjacent whenever 
the corresponding edges of $G$ are. 

The clustering coefficient for the node $a$ in the transformed network can be written by using Eq. (1) as: 
\begin{equation}
C_a (k) \simeq \frac{2[(k^\prime-1)\cdot(k^\prime-2)/2 + (k^{\prime\prime}-1)\cdot(k^{\prime\prime}-2)/2]}{(k^\prime-1 + k^{\prime\prime}-1)
(k^\prime -1 + k^{\prime\prime} - 1 - 1)},
\end{equation}
where $k=k^\prime + k^{\prime\prime} - 2$, because the edge $a$ vanishes in the graph $L(G)$. This equation 
ignores cases where edges in the graph $G$, $b$ and $b^\prime$ for example, have a common node as endpoint (i.e., existence of triangles or 
{\it loops} in Fig. 2(c)). However, we can quantify these cases by using a new parameter $l$. As we can see in Fig. 2(c)-(d), edges with 
one common node as
endpoint in the graph $G$ means one additional edge in the graph $L(G)$. This additional edge in $L(G)$ connects two 
neighbors of node $a$. By following definition of Eq. (1), it means
that $n_a$ increases its value by one unit. We can consider these cases by increasing one unit the parameter $l$ for 
each common node as endpoint of two edges in the graph $G$ (for example, $l=1$ means one common node). We write Eq. (4) 
after introducing the parameter $l$ as \cite{note2}: 

\begin{figure}[h]
\centerline{\protect
\hbox{
\psfig{file=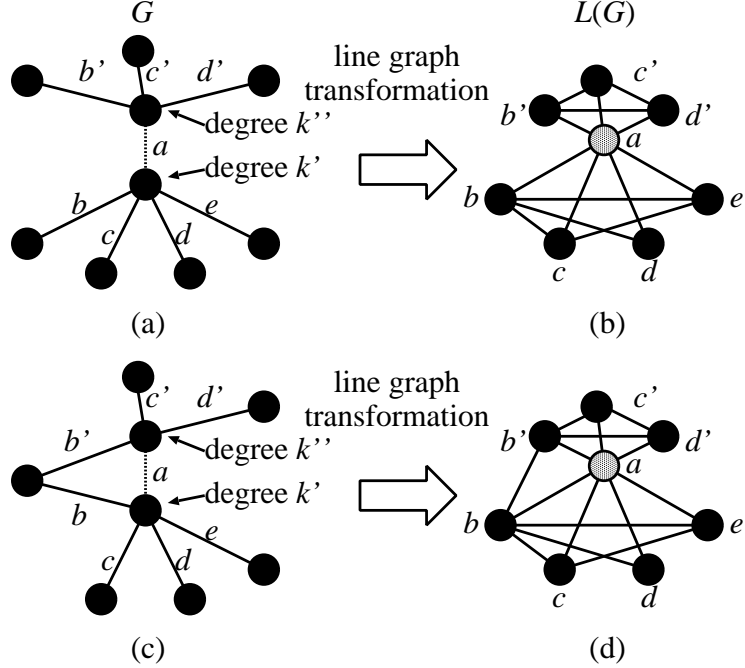,height=9.0cm,angle=0}}}
\caption{\small{ (a) Graph $G$ with two hubs with degree $k^\prime$ and $k^{\prime\prime}$ connected by edge $a$. (b) The corresponding 
line graph 
$L(G)$ after the line graph transformation is done. (c) Graph $G$ where edges $b$ and $b^\prime$ have a common node as endpoint. (d) Line 
graph of (c). It is worth noticing that (d) has only one more edge than (b). Hence, (d) has one more triangle that go through
node $a$ than (b).}}
\end{figure}

\begin{equation}
C_a (k) = \frac{2[(k^\prime-1)\cdot(k^\prime-2)/2 + (k^{\prime\prime}-1)\cdot(k^{\prime\prime}-2)/2 + l]}{(k^\prime-1 + k^{\prime\prime}-1)
(k^\prime -1 + k^{\prime\prime} - 1 - 1)},
\end{equation}
where if $l=0$ means that there are not {\it loops} and we recover Eq. (4). It should be noted that $l$ always contributes to increasing the value of $C_a(k)$ and $C_a(k)\leq 1$ always holds from the definition. In order to study the limits of Eq. (5) we consider the following 
two cases:

\begin{itemize}
\item $a)$ $k^\prime$=$k^{\prime\prime}$: We analyse the case where both degrees have the same value. We also consider the cases 
when $l=0$ and $l\neq 0$ in order to study the effect of triangles. We show the results in Fig. 3. For large $k^\prime$, Eq. (5) goes 
asymptotically 
to $1/2$ for $l=0$ and $l\neq 0$. We also see that for $k^\prime \geq 25$, all lines are very close
to $1/2$. For low $k^\prime$ and $l=0$, $C_a(k)$ takes values from 0.33 ($k^\prime$ =3) to 0.48 ($k^\prime$ =20). Hence, we see 
in Fig. 3 that
higher values of $l$ (more triangles) increase the values of $C_a(k)$. 

\item $b)$ $k^{\prime\prime} = constant$, $k^\prime >> k^{\prime\prime}$: We plot in Fig. 4 three cases. $k^{\prime\prime}$ is fixed with 
constant values as $k^{\prime\prime} =5$ (black),
 $k^{\prime\prime} =10$ (red), $k^{\prime\prime} =20$ (blue) and $k^\prime$ is a free parameter. We see that $C_a(k)$ 
 approaches to 1 when $k^\prime$ takes large values. For low $k^\prime$, the case $k^{\prime\prime}$ = 5 shows a 
 minimum with a few values of
 $k^\prime$ below 1/2. As we can see with dotted and dash-dotted lines in Fig. 4, the presence of triangles ($l\neq 0$) increases 
 the value of $C_a(k)$. Finally, for
 $k^{\prime\prime}$ =10 and $k^{\prime\prime}$ =20, we see that only a few values of $C_a(k)$ are slightly below 1/2 for low $k^\prime$.
 This analysis is complemented by calculating the minimun value of $C_a(k)$ analitically as: $\frac{\partial{C(k)}}{\partial k^{\prime}} = 0.$
 The value of $k^{\prime}$, where the function $C_a(k)$ takes the minimum value, is given by \cite{note3}: 
 
\begin{equation}
k^\prime=\frac{-1+l+k^{\prime\prime}+
\sqrt{2+3l+l^2-7k^{\prime\prime}-5lk^{\prime\prime}+9(k^{\prime\prime})^2+2l(k^{\prime\prime})^2-5(k^{\prime\prime})^3+(k^{\prime\prime})^4}}{-1+k^{\prime\prime}}
\end{equation}

 By substituting this equation into Eq. (5), it is possible to calculate the minimum value of $C_a(k)$ for each configuration of $l$ and
 $k^{\prime\prime}$.

\end{itemize}

 From these two cases, we can conclude that for hubs (i.e., those nodes with high degree ($k^\prime$ and $k^{\prime\prime}$ $>>1$)) and 
 for highly clustered networks (many triangles $l>>1$), the values of $C_a(k)$ in the transformed network are between
 around [$\frac{1}{2}, 1$].

\begin{figure}[h]
\centerline{\protect
\hbox{
\psfig{file=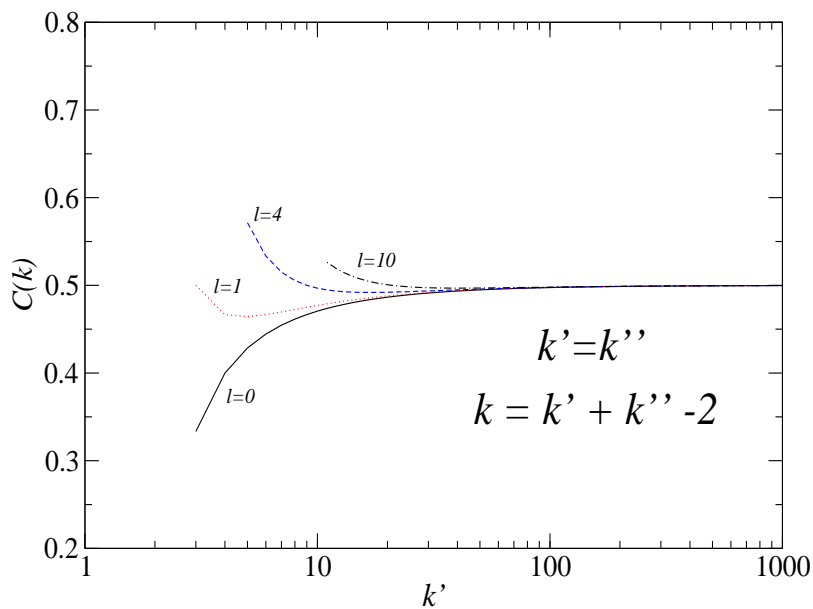,height=10.0cm,width=12.0cm,angle=-90}}}
\caption{\small{Values of $C_a(k)$ from Eq. (5) calculated by taking $k^\prime$=$k^{\prime\prime}$. Number of 
common nodes as endpoint of two edges (triangles) are indicated by the parameter $l$. The degree of transformed nodes is 
$k=k^\prime + k^{\prime\prime} - 2$ because the edge $a$ vanishes in the graph $L(G)$.}}
\end{figure}

\begin{figure}[h]
\centerline{\protect
\hbox{
\psfig{file=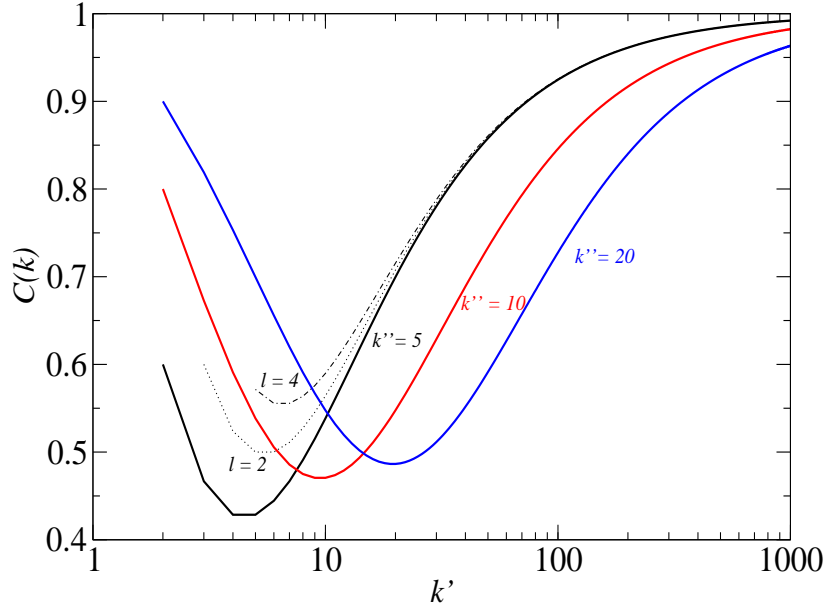,height=10.0cm,width=12.0cm,angle=-90}}}
\caption{\small{Values of $C_a(k)$ from Eq. (5) calculated by taking $k^{\prime\prime}$ with 
constant values as $k^{\prime\prime} =5$ (black line),
 $k^{\prime\prime} =10$ (red), $k^{\prime\prime} =20$ (blue) and $k^\prime$ as a free parameter. Dotted and dash-dotted 
 lines show the presence of triangles ($l\neq 0$). Triangles increase the value of $C_a(k)$.}}
\end{figure}




To calculate the distribution of $C(k)$ in the transformed space $(C^T(k))$ we introduce the concept of 
assortativity. By assortative (disassortative) 
mixing in networks we understand the 
preference for nodes with high degree to connect to other 
high (low) degree nodes \cite{new_newman}. By following {\it Newman} \cite{new_newman}, we define the probability distribution to choose a randomly
edge with two nodes at either end with degrees $k^\prime$ and $k^{\prime\prime}$ as $e_{k^\prime\,k^{\prime\prime}}$. We also assume that the
nodes of the initial network are following a power-law distribution $k^{-\gamma}$ and have no assortative mixing.  
Under these assumptions, the probability distribution $e_{k^\prime\,k^{\prime\prime}}$ of edges that link together nodes with 
 degree $k^\prime + k^{\prime\prime}$ can be written as:

\begin{equation}
e_{k^\prime\,k^{\prime\prime}}=\frac{k^{\prime -\gamma+1}}{\sum_{k^\prime} k^\prime}\frac{k^{\prime\prime -\gamma+1}}
{\sum_{k^{\prime\prime}} k^{\prime\prime}}.
\end{equation} 

We make a convolution between Eq. (4) and Eq. (7), by summing for all the possible degrees of the two nodes at either end of 
edges ($k^\prime,  k^{\prime\prime})$, which can generate
transformed nodes with degree $k=k^\prime + k^{\prime\prime} - 2$. Thus, we obtain:


\begin{equation}
C^T(k) \simeq \frac{\sum_{k=k^\prime + k^{\prime\prime} -2} k^{\prime\, -\gamma+1} \cdot 
k^{\prime\prime\, -\gamma+1} \cdot C_a(k)}{\sum_{k=k^\prime+k^{\prime\prime} -2} k^{\prime\, -\gamma+1} \cdot 
k^{\prime\prime\, -\gamma+1}}.
\end{equation}
According to the structure of $C^T(k)$ 
and the behavior of $C_a(k)$ exposed above, $C^T(k)$ will grow smoothly for large $k$, i.e., scaling weakly 
with the node degree $k$. We have calculated numerically 
this expression and the results are discussed later in Fig. 6.

We have also calculated the analytical expression for $C(N)$, and we have found that $C(N)$ has a size-independent behavior 
before and after the line graph transformation is done. We can write the number of nodes with degree $k$ as:
\begin{equation}
N_k \propto \frac{N\cdot k^{-\gamma}}{\sum_{k=1}^\infty k^{-\gamma}}
\end{equation}
and we assume that $C(k) = A\cdot k^{-\alpha}$, where $A$ is a constant. This constant changes
when we consider hierarchical networks with different number of nodes in the initial cluster \cite{jerar1}. But it seems natural 
because in that case the degree distribution 
$P(k)\sim k^{-\gamma}$ of the network
also changes. For $C(N)$ before the transformation we can write \cite{note4}: 

\begin{equation}
C(N) = \frac{1}{N} \sum_{i=1}^{N} C_i \simeq \frac{1}{N}{\sum_{k=2}^\infty N_k C(k)} = A \frac{\sum_{k=2}^\infty k^{-(\alpha+\gamma)}}
{\sum_{k=1}^\infty
k^{-\gamma}}.
\end{equation}
Furthermore, if we use the RSMOB model (explained in next section), Eq. (10) takes the form:

\begin{equation}
C(N) \simeq A^\prime\sum_{j=1}^{\log_m N}\frac{[(m-1)^j]^{-\gamma^{\prime}}[(m-1)^{-\alpha}]}{[(m-1)^j]^{-\gamma^{\prime}}},
\end{equation}
where $j$ is the number of iterations,  $m$ is the number of nodes in the initial fully connected cluster and $A^\prime$ is a 
constant adjusted so that $C(N)$=1 holds for $j=1$. 
The upper limit of the summation $\log_m N$ is obtained by means of the expression $m^j=N$, which gives the total number of
nodes in the network and  $\gamma^{\prime}$ = 
$\frac{\ln m}{\ln (m-1)}$ denotes the exponent of the power-law distribution of hubs in the RSMOB model \cite{note5}. 
The 
approximately equal symbol indicates that Eq. (11) is valid for hub nodes, and non-hub nodes are not considered.

By using these equations, we will see later (Tables 1 and 2) that $C(N)$ converges to a constant. In order to calculate $C(N)$ after the line graph transformation 
is applied ($C^T(N)$), we make the substitution $C(k)\rightarrow C^T(k)$ in Eq. (10). As from Eq. (8) we have seen that $C^T(k)$ is almost
constant, we can conclude that $C^T(N)$ also has a constant behavior and it is almost independent with network size.

While the scaling law of $C(k)\sim k^{-1}$ was proved mathematically in \cite{jerar1}, here we have 
obtained the analytical expressions of $C^T(k)$, $C(N)$ and $C^T(N)$. 

\section{Results and Discussion}

The RSMOB model \cite{jerar1} is able to reproduce the main topological features of 
the metabolic network. We follow the method described in \cite{jerar1} and generate a hierarchical network. Then, we apply 
the line graph transformation 
to that network. 

Fig. 5
illustrates the hierarchical network generated by the RSMOB model. The network is made of densely linked 5-node modules 
(it is worth noticing that the number of nodes in the
initial module can
be different than 5) that are assembled into larger 25-node modules (iteration n=1, $5^2 = 25$ nodes).
In the next step four replicas are created and the peripheral nodes are connected again to produce 125-node modules (iteration n=2, $5^3=125$
nodes). This process
can be repeated indefinitely \cite{jerar1, jerar2}. 

\OMIT{
\begin{figure}[h]
\centerline{\protect
\hbox{
\psfig{file=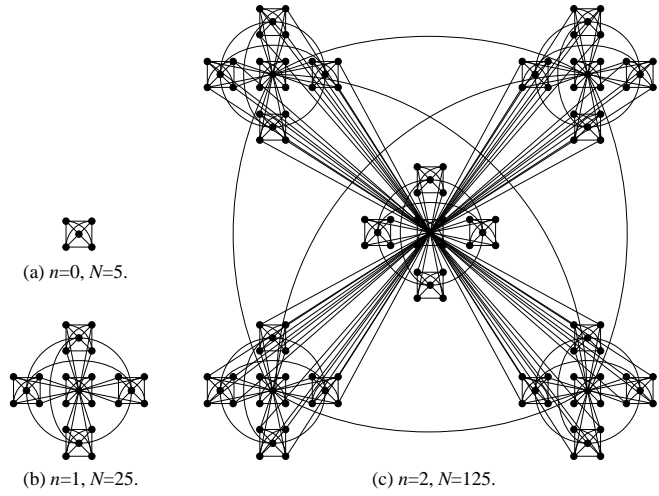,height=6.5cm,angle=0}}}
\caption{\small{Hierarchical network generated by using the RSMOB model \cite{jerar2}. Starting from a fully connected cluster of 5 nodes, 4
identical replicas are created, obtaining a network of N=25 nodes in the first iteration n=1 ($5^2=25$ nodes). The model differs slightly from
that one from \cite{jerar2} because we have linked to each other the central hubs of the replicas. This process can repeated indefinitely. We note that
the initial number of nodes can be different than 5.}}
\end{figure}
}

To evaluate $C(k)$, we have constructed three hierarchical networks with 3, 4, and 5 
initial number of nodes. These networks were generated up to 7 (6561), 5 (4096), and 4 (3125) iterations (nodes), repectively. Once 
we have constructed these 
three networks, we apply the line graph
transformation to them, and we calculate the $C^T(k)$ clustering coefficient for the transformed networks. In Fig. 6(a) we show the 
results of the
clustering coefficient of the transformed network. Circles, triangles and squares indicate the values of $C(k)$ for the transformed network
with 3, 4, and 5 initial nodes, respectively. In Fig. 6(a) we also plot with continuous lines the values of $C^T(k)$ obtained from Eq. (8). From top 
to bottom the lines correspond to the networks of 3, 4 and 5 initial nodes, respectively. 
In Fig. 6(a), we see that the lines show an acceptable agreement with the overall tendency of data generated by the transformed 
network. In Fig. 6(b), we 
see that the results from theoretical calculation of $C^T(k)$ via Eq. (8) (lines) are in good agreement
 with the experimental data (diamonds) 
 from the KEGG database
\cite{KEGG}. The only disagreement comes at $k=2$. This is easy to understand
because in the hierarchical model depicted in Fig. 5, we can only find $C(k=2)=1$ for 3 initial nodes by construction of
the network. However, in real networks, we could find nodes which have only two neighbors and, in some cases, 
these neighbors could be connected. In these cases the clustering coefficient takes value one.


\OMIT{
\begin{figure}[h]
\subfigure{\includegraphics[height=8.5cm,width=10.0cm,angle=-90,keepaspectratio]{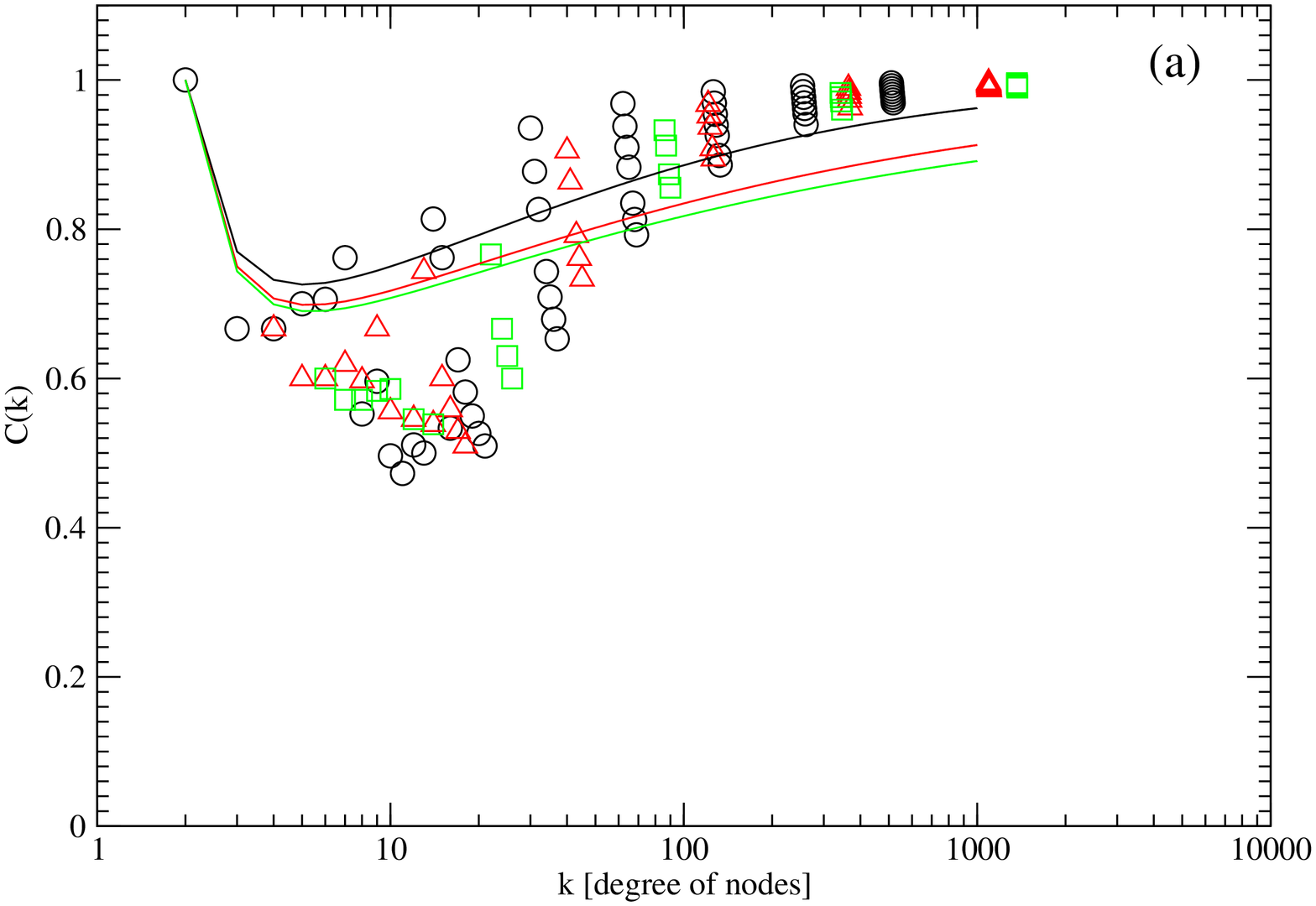}}
\subfigure{\includegraphics[height=8.5cm,width=10.0cm,angle=-90,keepaspectratio]{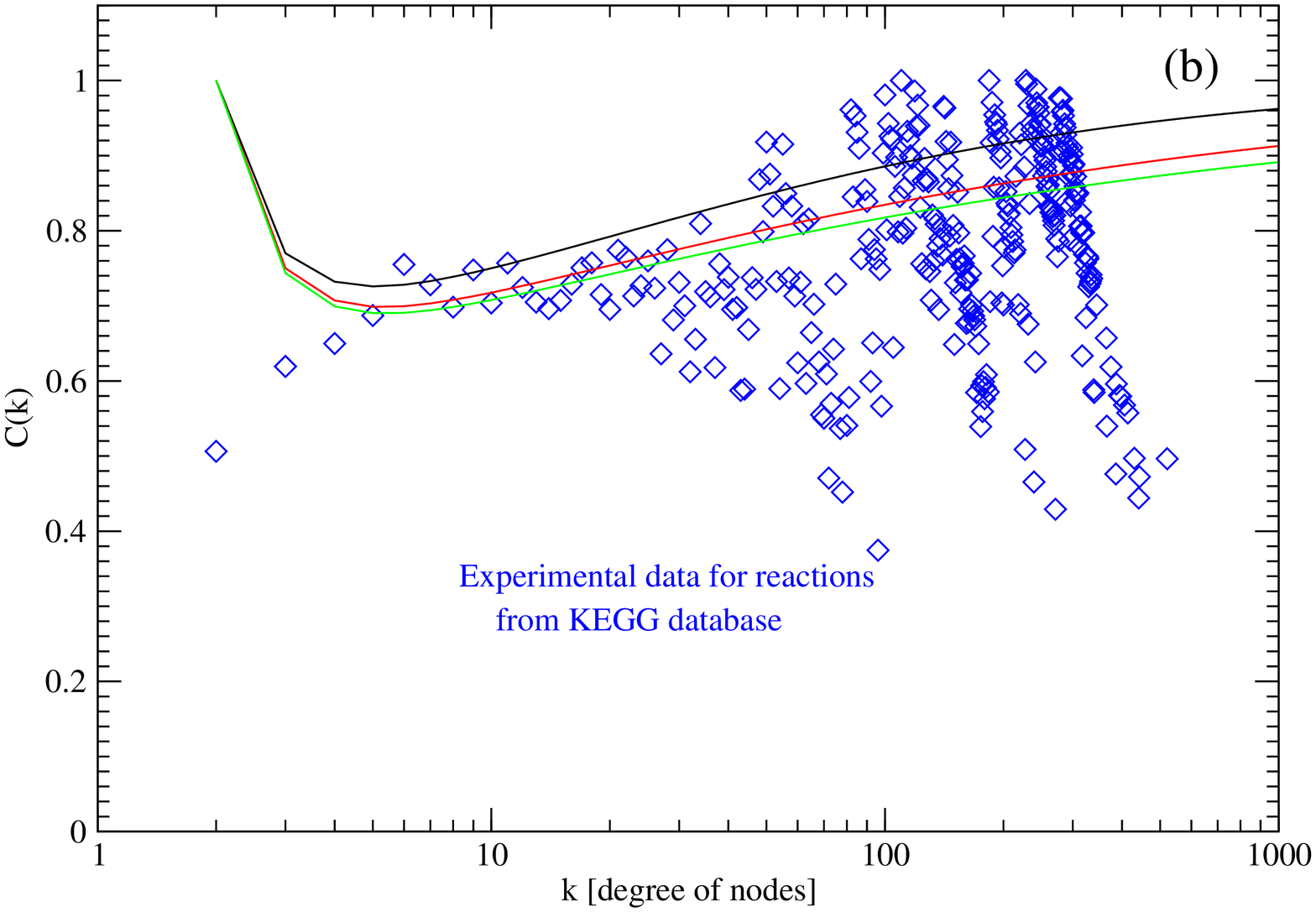}}
\caption{\small{a) We plot the results of the hierarchical model for $C^T(k)$ for different configurations. 3 initial nodes and up to 7 iterations (circles), 4
initial nodes and up to 5 iterations (triangles), 5 initial nodes and up to 4 iterations (squares). From top to bottom (3 initial nodes (black), 
4 initial nodes (red), 5 initial nodes (green)), we show with lines the results of $C^T(k)$ obtained by means of Eq. (8). b) The lines
have the same meaning as
before and the diamonds correspond to the experimental data for reactions from the KEGG database \cite{KEGG}. Experimental data involves 
163 organisms.}}
\end{figure}
}


In Fig. 7, we show the results for $C(k)$ after the line graph transformation is applied to the hierarchical network 
generated by 4 initial nodes 
and up to 5 iterations. The results are shown with empty triangles (red) and fitted to the dashed line. We 
see that  $C(k)\sim k^{-1.1}$ changed into $C^T(k)\sim k^{0.08}$. We also see that the line 
graph transformation increases the average of the clustering value of the transformed
network. These theoretical results were compared with the experimental data from KEGG \cite{KEGG}, finding 
a good agreement, and supporting the result of a degree-independent clustering coefficient $C^T(k)$ for the reaction network. 


\OMIT{
\begin{figure}[h]
\centerline{\protect
\hbox{
\psfig{file=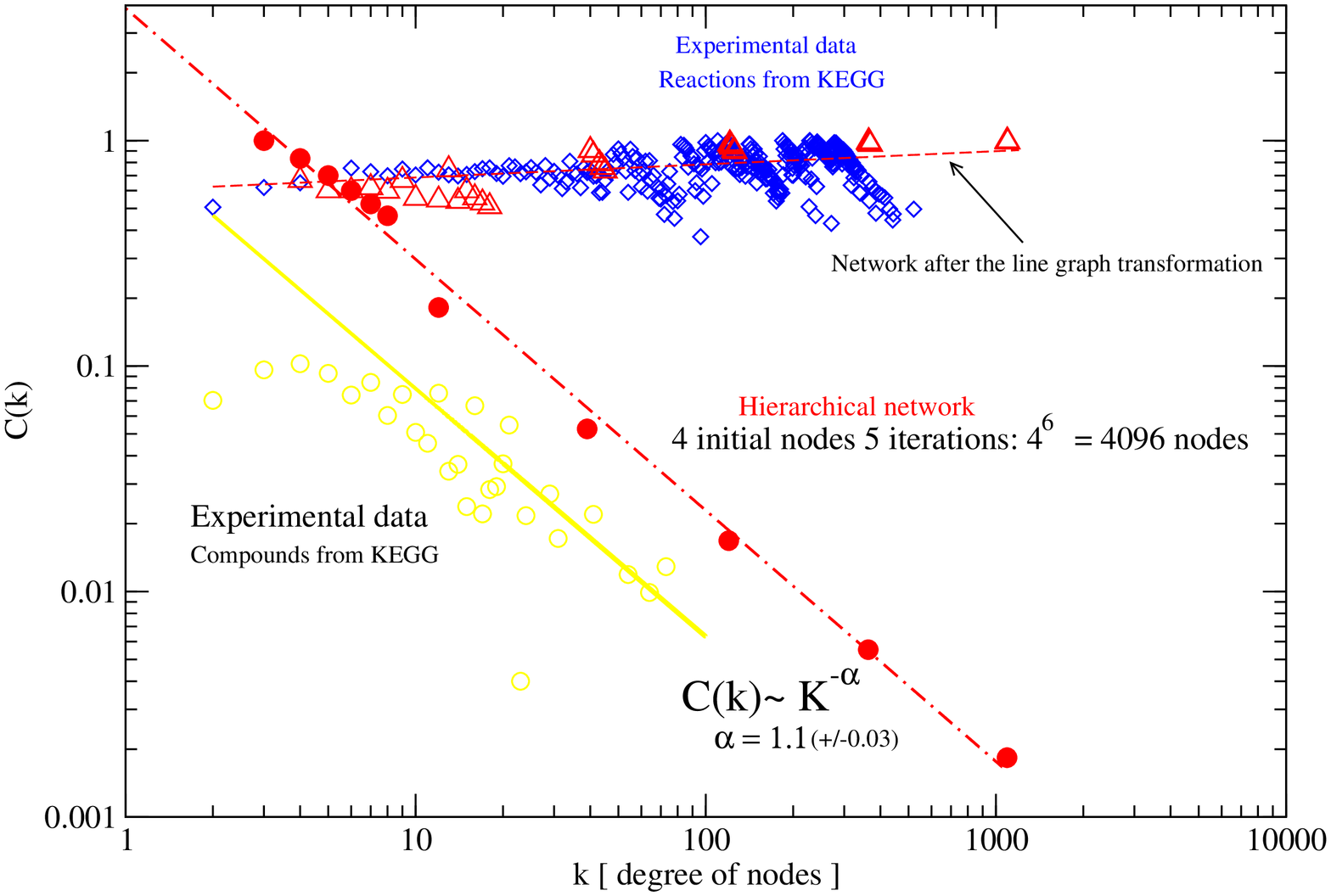,height=9.0cm,angle=-90}}}
\caption{\small{Full circles (red) and dot-dashed line (red): $C(k)$ evaluated with the hierarchical network. Empty triangles (red) and 
dashed line (red): $C(k)$ after the line graph transformation is done over the hierarchical network ($C^T(k)$). 
Diamonds (blue): $C(k)$ of reactions data from the KEGG database \cite{KEGG}. Empty circles (yellow) and continuous line: $C(k)$ of compounds data from KEGG. Hierarchical model with 4 initial nodes and 5 iterations.}}
\end{figure}
}

For $C(N)$ we have evaluated Eq. (10) for 3 different configurations. We have considered 3 initial nodes, 4 and 5
initial nodes nodes up to 7, 5 and 4 iterations, respectively. As it is explained in \cite{jerar1,jerar2}, $C(N)$ 
approaches asymptotically to a constant value, being independent of the size of the network.
The asymptotic value depends on the initial number of nodes. We calculated the values of $\gamma$ corresponding to the 
degree distrution $P(k)\sim k^{-\gamma}$ for each network, and the
related constant $A$, which appears in Eq. (10). We show in Table 1 the values of these parameters and the results 
of $C(N)$ obtained by
Eq. (10). These values, as it can be seen in Fig. 8(a), are below
the asymptotic values of $\sim 0.66$ (circles) and $\sim 0.74$ (triangles) obtained by using the RSMOB model. However, we 
have found an explanation for this result. In Fig. 7, the full circles
at the top of the dash-dotted line correspond to non-hubs nodes. We have
checked that these nodes do not follow a power-law, hence the value of $C(k)$ is being overestimated by the
scaling dependence $k^{-1}$ and it
provides a
larger value of $C(N)$. In \cite{jerar1}, the values of $C(N)$ from hierarchical model were 
compared with the experimental values of 43 organisms. The values of $C(N)$ for each organism were around 0.15 - 0.25. By using 
the KEGG database we have evaluated the 
experimental value $C(N)$ for 163 organisms and we obtained an average value of 0.08. 



We show in Fig. 8(a) the values of $C(N)$ calculated for networks generated by 3 initial nodes (circles) and 4 initial nodes 
(triangles) by using the RSMOB model. We see that
$C(N)$ approaches asymptotically to constant values around $\sim 0.66$ (circles) and $\sim 0.74$ (triangles), being independent 
of the size of the network. Once the line graph transformation 
is applied, we see that the
corresponding values of $C^T(N)$ also approach asymptotically to constant values. Hence, $C^T(N)$
also is size-independent for large $N$ (empty circles
and triangles). In addition, we have averaged the experimental value of the 
clustering coefficient for reactions of 163 organisms found in KEGG
database and we have obtained the value of $C^T(N)$=0.74. We see that the experimental value $C^T(N)$  for reactions 
is in good agreement with the asymptotic values obtained by the transformed network (empty triangles and circles).

Furthemore, we have also calculated $C(N)$ by using Eq. (11). This equation should reproduce the results of 
$C(N)$ calculated by using the RSMOB model (dark circles and triangles in Fig. 8(a)). In Fig. 8(b), we see that 
the results are qualitively similar to those shown in Fig. 8(a) (dark circles and triangles).

We remark
that the theoretical analysis of $C(N)$ and $C^T(N)$
done here has also been useful to prove that they are independent of network size.


\begin{table}
\begin{center}
\begin{tabular}{|c|c|c|c|c|} \hline
$m$ initial ( total ) nodes & $\gamma$  & $\alpha$  &  $A$ 
  & $C(N)$ (Eq.(10))\\ \hline
3 (6561) & 2.58 & 1.1 & 2.34 & 0.20  \\ \hline
4 (4096) & 2.26 & 1.1 & 3.68 & 0.36  \\ \hline
5 (3125) & 2.16 & 1.1 & 5.18 & 0.54  \\ \hline
\end{tabular}
\end{center}
\caption{\small{Results of $C(N)$ evaluated by using Eq. (10) and the needed parameters in that calculation for 3 different setups:
$\gamma= 1 + \gamma^\prime$, where $\gamma^\prime$= $\frac{\ln m}{\ln(m-1)}$ ($P(k)\sim k^{-\gamma}$), $\alpha$ ($C(k)\sim k^{-\alpha}$), $A$ 
($C(k)=A\cdot k^{-\alpha}$). Eq. (10) is a general expression of $C(N)$ .
}}
\end{table}

\begin{table}
\begin{center}
\begin{tabular}{|c|c|c|c|c|} \hline
$m$ initial ( total ) nodes & $\gamma^\prime$  & $\alpha$  & $C(N)$ (Eq.(11))\\ \hline
3 (6561) & 1.58 & 1.1 & 0.78  \\ \hline
4 (4096) & 1.26 & 1.1 & 0.81 \\ \hline
5 (3125) & 1.16 & 1.1 & 0.83  \\ \hline
\end{tabular}
\end{center}
\caption{\small{Results of $C(N)$ evaluated by using Eq. (11) for 3 different setups. The exponent of the power-law distribution of hubs is
given by $\gamma^\prime$= $\frac{\ln m}{\ln(m-1)}$. The parameter $\alpha$ has same meaning as in Table 1.  We also notice that in Eq.
(11), $A^\prime$ is adjusted so that $C(N)$=1 holds for $j=1$. Eq. (11) is the particular expression of $C(N)$
applied to the RSMOB model.}}
\end{table}

\OMIT{
\begin{figure}[h]
\subfigure{\includegraphics[height=8.5cm,width=10.0cm,angle=-90,keepaspectratio]{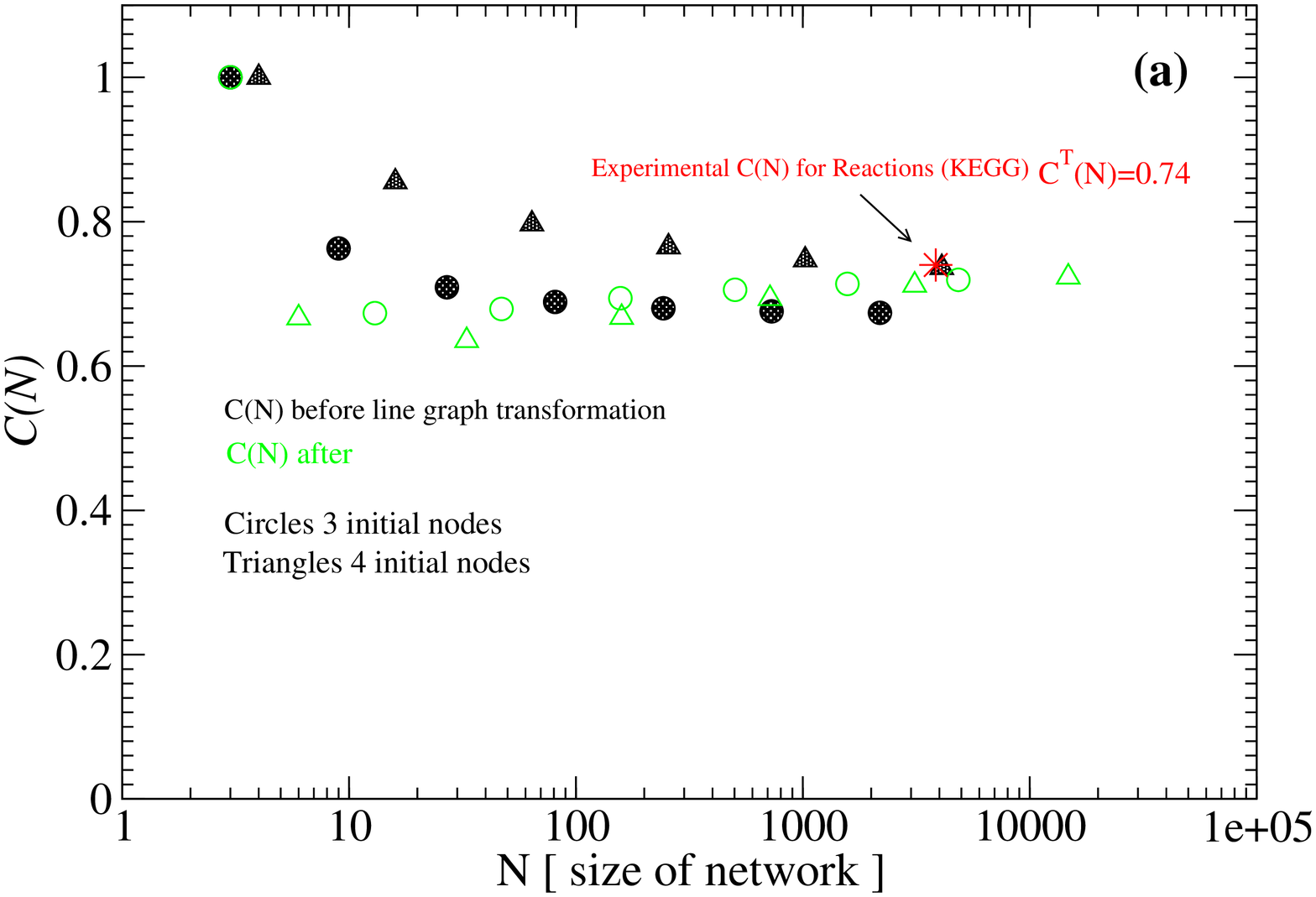}}
\subfigure{\includegraphics[height=8.5cm,width=10.0cm,angle=-90,keepaspectratio]{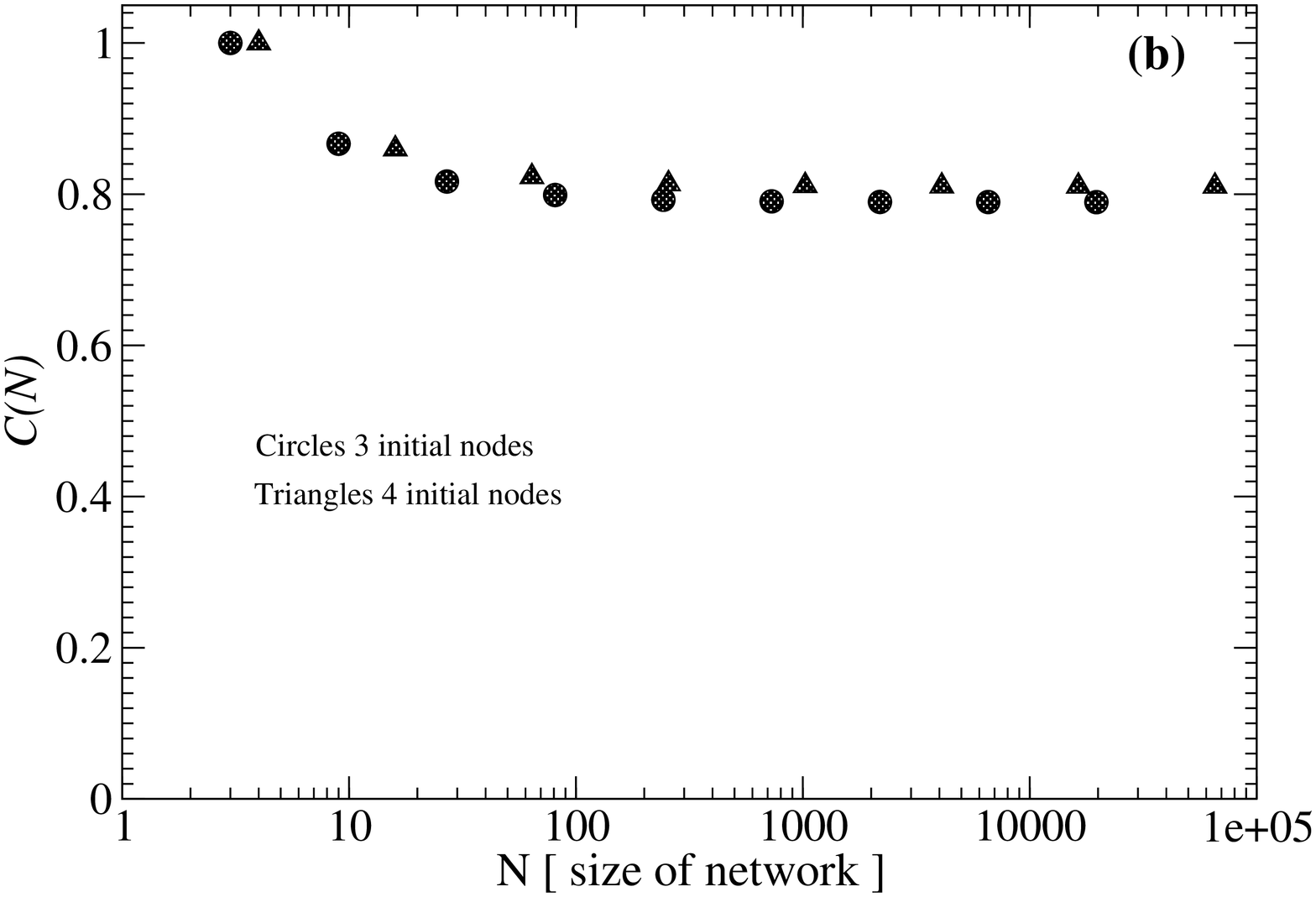}}
\caption{\small{a) Dark (black): $C(N)$ is calculated by using the hierarchical network. Light (green):
$C^T(N)$ ($C(N)$ after the line graph transformation is applied to 
the hierarhical network). Circles (3 initial nodes), Triangles (4 initial nodes). Star (red): Experimental $C^T(N)$ for
reactions from the KEGG database \cite{KEGG}.  b) $C(N)$ is calculated by using Eq. (11). The results show a good agreement and similar tendency
to those shown in Fig. 8(a) (dark circles and triangles). }}
\end{figure}
}


Finally, in Fig. 9 we plot the hierarchical network (left) and the transformed network (right) 
by using the graph drawing tool {\it Pajek} \cite{pajek}. We 
see the high degree of compactness of the transformed network. It could be related to the concept of robustness of a network. It means 
that by removing one node randomly 
from the reaction network depicted in
the Fig. 9, the normal
behavior of the cell might be preserved by finding an alternative path (reaction) 
to complete the task. This fact could be a consequence of the high degree of clustering and
connectivity between the nodes in the transformed network.

\OMIT{
\begin{figure}[h]
\subfigure{\includegraphics[height=4.0cm,angle=-90,keepaspectratio]{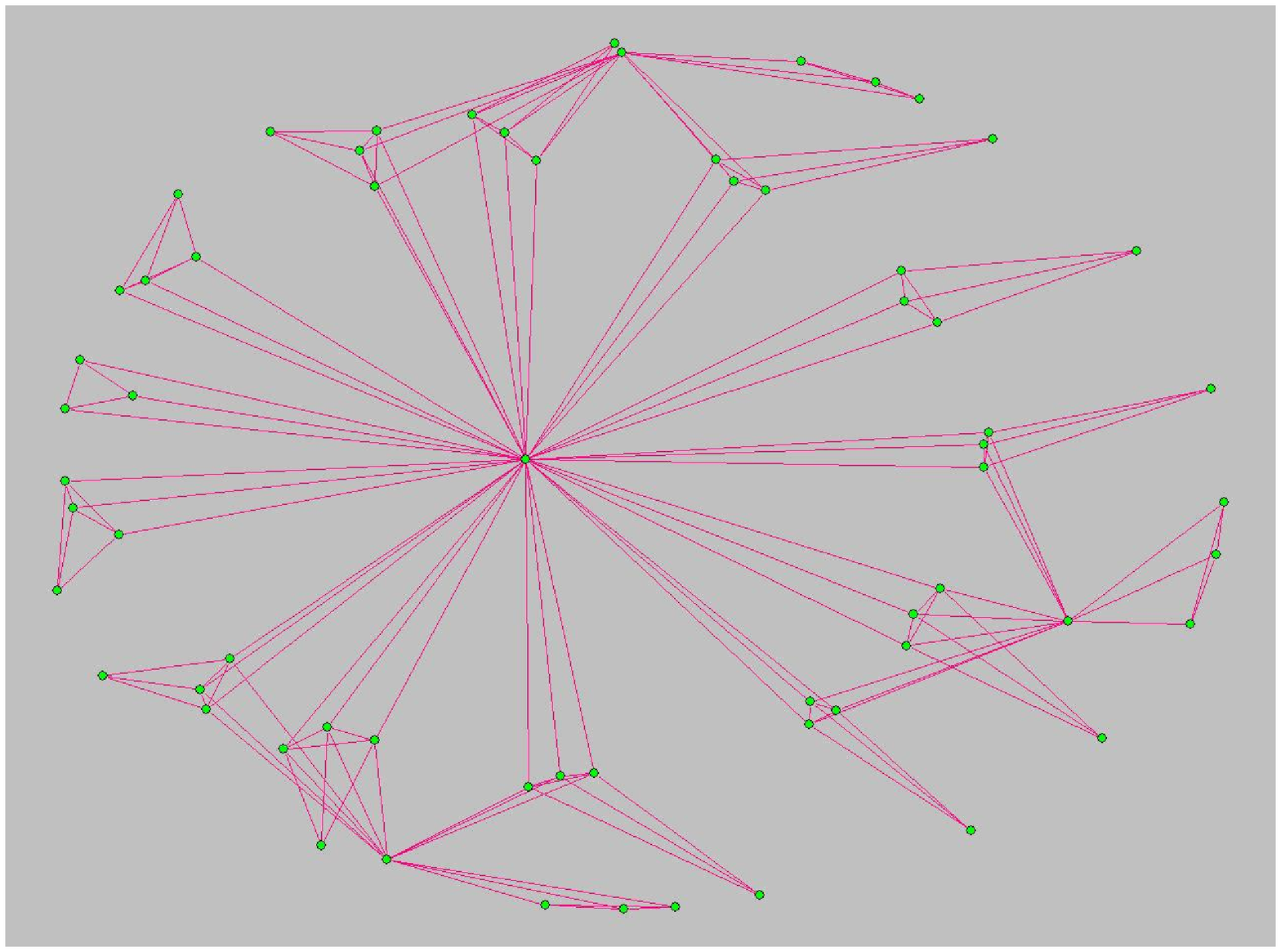}}
\subfigure{\includegraphics[height=4.0cm,angle=-90,keepaspectratio]{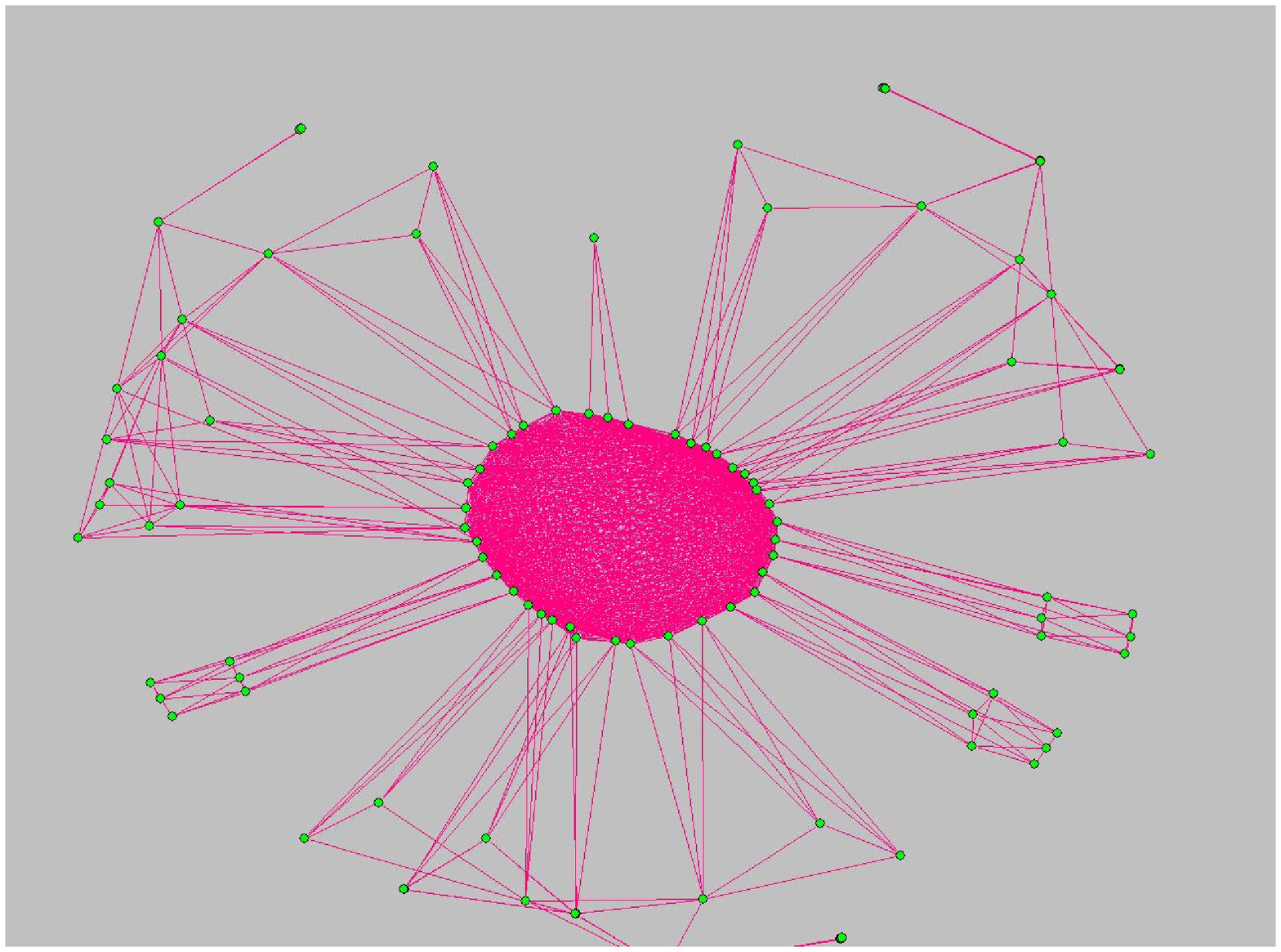}}
\caption{\small{Left: Hierarchical network generated by using the model of ref. \cite{jerar2} with 4-node modules and up to 2 
iterations. Right: Network 
after the line graph transformation. We see a huge interlinked cluster in the center of figure, which generates the degree-independent
clustering coefficient $C(k)$ (it scales weakly as $C(k)\sim k^{0.08}$ ). }}
\end{figure}
}

\begin{table}
\begin{center}
\begin{tabular}{|c|c|c|c|} \hline
Func. & Definition & Dependence ${\it before}$ & Dependence ${\it after}$ \\ \hline
$P(k)$ & $N_k/N$  & $k^{-\gamma}$ & $k^{-\gamma + 1\, \star}$ \\ \hline
$C_i(k)$ & $2n / [k_i(k_i-1)]$ & ${k^{-1.1}}$ & $k^{0.08\, \dagger}$  \\ \hline
$C(N)$ & $[\sum_i C_i(k)] / N$ & size-independent$^\dagger$ & size-independent$^\dagger$ \\ \hline
\end{tabular}
\end{center}
\caption{\small{Definitions of functions and their values before and after the line graph transformation is applied to
the hierarchical network. $N_k$: number of nodes of degree $k$. The $\dagger$ symbol means that these dependences were analyzed in the present work,
while the $\star$ symbol means that it was studied in our previous work \cite{nacher}.}}
\end{table}
At the end of this section, it is convenient to summarize our findings in Table 3.
We can see in Table 3, with $\dagger$ and $\star$ symbols, the functions studied analytically and evaluated by us. We see 
that central properties of networks were studied by using the line graph transformation technique, which suggests the
effectiveness of the method.


\section{Conclusions}

We have studied here the
clustering coefficients $C(k)$ and $C(N)$ of the reaction network 
by applying the line graph transformation to a hierarchical network. This hierarchical network was generated by using the RSMOB model, which
reproduces properly the topological features of the metabolic network, in particular the compound network. Our results indicate 
that by applying the line graph transformation to the hierarchical network, it is possible to extract  
topological properties of the reaction network, which is embedded in the metabolic network. The RSMOB model stores the adequate information
of the reaction network and the line graph transformation is one useful technique to evoke it.

While $C(k)$ scales as $k^{-1.1}$ for the initial 
hierarchical network (compound network), we find $C(k)\sim k^{0.08}$ for the transformed network (reaction network). This 
theoretical prediction was compared with the experimental data from the KEGG database, finding a good agreement. Our results indicate that
the reaction network is a degree-independent clustering network. Furthermore, the weak scaling of $C(k)$ for the reaction network suggests us that this network has not hierarchical organization. 

On the other hand, we have also conducted an analytical derivation for the clustering coefficients $C(k)$ and $C(N)$. Expressions for these coefficients were
calculated before and after the line graph transformation is applied to the hierarchical network. The agreement obtained by using 
these expressions was found acceptable, and consequently, they could be useful for further analyses.

The line graph transformation has recently been applied on metabolic networks \cite{nacher} to study the scale-free topology of the reaction network, and
on the protein-protein interaction network to detect functional clusters \cite{protein}. The work done here is another important application of this interesting technique.

\vspace{0.5cm}
\noindent


{\bf{Acknowledgements}}

One of us, J.C.N. wishes to thank the support of the postdoctoral programme in Bioinformatics 
with funding from the Ministry of Education, Culture, Sports, Science and Technology (MEXT) of Japan. This work 
was partially supported by Grant-in-Aid for Scientific Research on Priority Areas (C) ``Genome Information Science''
from MEXT.



\begin{thebibliography}{99}
\bibitem{note1} We notice that we have not considered water and ATP metabolites as edges.
\bibitem{note2} Though it is more realistic to consider the parameter $l$ as a function of $k^\prime$ and
$k^{\prime \prime}$, we have considered $l$ as an independent parameter. However, this simplification does not affect 
the qualitative features of our results.
\bibitem{note3} Positive solution of the square root is written.
\bibitem{note4} We note that the summation in the denominator begins with $k=1$  because we renormalize over all 
the probability distribution.
\bibitem{note5} We must notice
that in a hierarchical network, the number of nodes with different degree $k$
is scarce, therefore the probability distribution of node degree is properly 
defined as $P(k)$=$(1/N_{tot})(N_k/\Delta k)$, where $N_k$ is the number of nodes 
with degree $k$, $N_{tot}$ is the total number of nodes,
and $\Delta k$ means that nodes with degree $k$ are binned into intervals.
In addition, we note that for the hierarchical model, $\Delta k$ changes linearly with $k$. Hence, the exponent of the power-law is given by
$\gamma = 1 + \gamma^\prime$, with $\gamma^\prime = \frac{\ln m}{\ln (m-1)}$ where $m$ is the number of nodes in the initial module.
\end{thebibliography}
\end{document}